\documentclass[%
 reprint,
 superscriptaddress,
 preprintnumbers,
 bibnotes,
 amsmath,amssymb,
 aps,
 nofootinbib,
longbibliography
]{revtex4-2}

 \usepackage[textsize=tiny
 ]{todonotes}

\usepackage[utf8]{inputenc}
\usepackage{graphicx}
\usepackage{tikz-feynman}
\usetikzlibrary{shapes.misc}
\usetikzlibrary{cd}
\usepackage{dcolumn}
\usepackage[caption=false]{subfig}
\usepackage{bm}
\usepackage{hyperref}
\hypersetup{
    colorlinks=true,
    linkcolor=Nblue,
    filecolor=black,      
    urlcolor=Nblue,
    citecolor=Nblue,
    breaklinks
}
\definecolor{Nblue}{RGB}{0,47,167}  

\usepackage{todonotes}
\usepackage{stackengine}

\usetikzlibrary{arrows}
\newcommand{\midarrow}{\tikz \draw [-{Stealth[length=2.2mm]}](0,0) -- (.1,0);}

\newcommand{\nint}{{n_{\text{int}}}}
\newcommand{\nISP}{{n_{\text{ISP}}}}
\newcommand{\LS}{{\text{LS}}}

\begin{document}

\title{Calabi-Yau meets Gravity: A Calabi-Yau three-fold at fifth post-Minkowskian order}

\author{Hjalte Frellesvig}
 \affiliation{Niels Bohr International Academy, Niels Bohr Institute, Copenhagen University, Blegdamsvej 17, 2100 Copenhagen \O{}, Denmark}
\author{Roger Morales}
 \affiliation{Niels Bohr International Academy, Niels Bohr Institute, Copenhagen University, Blegdamsvej 17, 2100 Copenhagen \O{}, Denmark}
 \affiliation{Mani L. Bhaumik Institute for Theoretical Physics, University of California at Los Angeles, Los Angeles, CA 90095, USA}
\author{Matthias Wilhelm}
 \affiliation{Niels Bohr International Academy, Niels Bohr Institute, Copenhagen University, Blegdamsvej 17, 2100 Copenhagen \O{}, Denmark}

\begin{abstract}
We study geometries occurring in Feynman integrals that contribute to the scattering of black holes in the post-Minkowskian expansion. These geometries become relevant to gravitational-wave production during the inspiraling phase of binary black hole mergers through the classical conservative potential.
At fourth post-Minkowskian order, a K3 surface is known to occur in a three-loop integral, leading to elliptic integrals in the result.
In this letter, we identify a Calabi-Yau three-fold in a four-loop integral, contributing at fifth post-Minkowskian order.
The presence of this Calabi-Yau geometry indicates that  completely new functions occur in the full analytical results at this order.
\end{abstract}

\maketitle

\section{Introduction}
Following the groundbreaking discovery of gravitational waves~\cite{LIGOScientific:2016aoc,LIGOScientific:2017vwq}, the inspiral and eventual merger of binary systems of compact astronomical objects such as black holes and neutron stars has become a key object of interest in many branches of physics. 
The upcoming third-generation gravitational-wave detectors will provide much more and higher-precision data, requiring equally high-precision theoretical predictions for its interpretation \cite{Berti:2022wzk,Buonanno:2022pgc}. 

Many complementary approaches for the theoretical description of these processes have been developed, ranging from numerical relativity~\cite{Pretorius:2005gq,Campanelli:2005dd,Baker:2005vv} to analytical approaches valid in various regions, such as post-Newtonian~\cite{Goldberger:2004jt,Blanchet:2013haa,Levi:2018nxp}, post-Minkowskian~\cite{Damour:2016gwp,Buonanno:2022pgc}, and self-force~\cite{Mino:1996nk,Quinn:1996am,Poisson:2011nh,Barack:2018yvs} expansions as well as the effective-one-body formalism \cite{Buonanno:1998gg,Buonanno:2000ef}.
 
The post-Minkowskian (PM) expansion treats the dynamics in the inspiraling phase 
perturbatively in Newton's constant $G$ while maintaining all orders in the velocity, thus accounting for relativistic effects. Since the dynamics of the bound system can also be related to the scattering problem~\cite{Damour:2016gwp}, this allows the use of Feynman diagrams and other methods from perturbative Quantum Field Theory (QFT) and scattering amplitudes~\cite{Bjerrum-Bohr:2018xdl,Cheung:2018wkq,Kosower:2018adc,Bern:2019nnu,KoemansCollado:2019ggb,Cristofoli:2019neg,Bern:2019crd,Kalin:2019rwq,Kalin:2019inp,Parra-Martinez:2020dzs,Kalin:2020mvi,Kalin:2020fhe,Mogull:2020sak,Herrmann:2021tct,Bern:2021dqo,Dlapa:2021npj,Bern:2021yeh,Dlapa:2021vgp,Cho:2021arx,Kalin:2022hph,Dlapa:2022lmu,Dlapa:2023hsl}, see refs.~\cite{Bjerrum-Bohr:2022blt,Buonanno:2022pgc} for an overview, while systematically taking the classical limit $\hbar \rightarrow 0$ to retain the classical pieces only. As in QFTs, higher precision thus requires the computation of Feynman integrals with more loops. In particular, the state-of-the-art computation for the gravitational two-body problem currently stands at three loops, corresponding to a 4PM correction for non-spinning black holes~\cite{Bern:2021dqo,Dlapa:2021npj,Bern:2021yeh,Dlapa:2021vgp,Dlapa:2022lmu,Dlapa:2023hsl}, as well as including spin-orbit~\cite{Jakobsen:2023ndj,Jakobsen:2023hig} and tidal effects~\cite{Jakobsen:2023pvx}.

With the objective of calculating Feynman integrals, one task is to characterize the space of functions to which they evaluate. Most Feynman integrals computed to date can be written in terms of multiple polylogarithms~\cite{Chen:1977oja,Goncharov:1995ifj}, which are iterated integrals over the Riemann sphere. 
However, at high loop orders and in cases with non-negligible masses or many physical scales, new special functions start to appear in QFT, involving integrals over non-trivial geometries.
These include integrals over elliptic curves, K3 surfaces and higher-dimensional Calabi-Yau manifolds; see ref.~\cite{Bourjaily:2022bwx} for a recent review. In particular, various $L$-loop families of Feynman integrals have been identified that involve Calabi-Yau manifolds of dimensions growing linearly with the loop order $L$ \cite{Broadhurst:1993mw,Bourjaily:2018ycu,Bourjaily:2018yfy,Bonisch:2021yfw,Broedel:2021zij,Duhr:2022pch,Lairez:2022zkj,Pogel:2022vat,Duhr:2022dxb,Cao:2023tpx}.

Up to two loops (3PM order), the results in the PM expansion are expressible in terms of polylogarithms. However, at three loops they contain products of complete elliptic integrals, which stem from a K3 surface~\cite{Bern:2021dqo,Dlapa:2021npj,Dlapa:2022wdu}. In this letter, we initiate an analysis beyond the current state of the art, finding that at four loops a new geometry appears -- a Calabi-Yau three-fold.  
This is the first instance that this type of geometry appears in integrals relevant for the scattering and inspiral of black holes, and it indicates that completely new functions are needed for the full analytical result at 5PM order.

In order to detect geometries in Feynman integrals, we use two complementary approaches: differential equations \cite{Kotikov:1990kg} and leading singularities \cite{Cachazo:2008vp}.
While leading singularities can, in principle, be calculated using any parametric representation of the Feynman integral as well as the loop-momentum-space representation, we find that a loop-by-loop Baikov representation~\cite{Baikov:1996iu,Frellesvig:2017aai} is particularly advantageous in the present case.
It allows us to show that the geometries occurring in many different PM Feynman diagrams are identical, and that further classes of diagrams contain only trivial geometries. 
This vastly reduces the number of different diagrams we need to consider for the purpose of detecting geometries, and even allows for a full classification, as we will show in upcoming work \cite{Frellesvig:toapp}.

The remainder of this letter is structured 
 as follows: In section~\ref{sec: setup}, we discuss which 
Feynman diagrams occur in the perturbative PM expansion of classical gravity. In section~\ref{sec: geometry}, we discuss our approach to detecting the geometry in the corresponding Feynman integrals; see our upcoming paper~\citep{Frellesvig:toapp} for further details. We present our results on relating geometries in different Feynman diagrams in sections~\ref{sec: results1} and the occurrence of a Calabi-Yau three-fold, depicted in fig.\ \ref{fig: kinematics}(b), in section \ref{sec: results2}. In section~\ref{sec: conclusion}, we present our conclusions and outlook. We include further details on the identification of the Calabi-Yau three-fold in appendix~\ref{sec: Baikov_Calabi_Yau}.

\section{Feynman diagrams for gravitational waves}
\label{sec: setup}

Our aim is to study the classical conservative dynamics for the two-body problem of two inspiraling, non-spinning black holes.\footnote{Spin effects can be included in our analysis, but they only modify the dressed vertices and thus the numerator of the integrals.
Similarly, the analogous description for neutron stars only affects the numerators.} For this, we will assume that the size of the bodies is much smaller than their distance, such that their internal degrees of freedom can be neglected. In the gravitational two-body problem, this condition is satisfied when the Schwarzschild radius $r_s \sim G m$ of each black hole is much smaller than the impact parameter $|b|$, in momentum space given as $|b|\sim 1/|q|$. Therefore, there is a small expansion parameter $r_s/|b| \sim G m |q| \ll 1$ inherent to the long-distance dynamics. This naturally defines a perturbative expansion that is compatible with the PM expansion of general relativity, corresponding to an $n$PM correction at order $G^n$. To study this problem, we will, in this work, furthermore use the modern scattering amplitudes-based approach to the PM expansion, where the two black holes are modeled by two massive scalars minimally coupled to gravity~\cite{Cheung:2018wkq}. In particular, we will closely follow the conventions of refs.~\cite{Parra-Martinez:2020dzs,Herrmann:2021tct,Bern:2021dqo,Bern:2021yeh}, but analogous results hold in all formulations of the PM expansion.

\begin{figure}[tb]
\centering
\subfloat[]{ \begin{tikzpicture}[baseline=(current bounding box.center),scale=1.4] 
	\node[] (a) at (0,0) {};
	\node[] (b) at (0,-1) {};
	\node[label=left:{$p_1$}] (p1) at ($(a)+(-1,0)$) {};
	\node[label=left:{$p_2$}] (p2) at ($(b)+(-1,0)$) {};
 	\node[] (p3) at ($(b)+(1,0)$) {};
 	\node[] (p4) at ($(a)+(1,0)$) {};
	\draw[line width=0.15mm, postaction={decorate}] (b.center) -- node[sloped, allow upside down, label={[xshift=0.75cm, yshift=0cm]$q$}] {\midarrow} (a.center);
	\draw[line width=0.5mm, postaction={decorate}] (p1.center) -- node[sloped, allow upside down] {\midarrow} (a.center);
	\draw[line width=0.5mm, postaction={decorate}] (a.center) -- node[sloped, allow upside down] {\midarrow} (p4.center);
	\draw[line width=0.5mm, postaction={decorate}] (p2.center) -- node[sloped, allow upside down] {\midarrow} (b.center);
	\draw[line width=0.5mm, postaction={decorate}] (b.center) -- node[sloped, allow upside down] {\midarrow} (p3.center);
\end{tikzpicture}}
\quad
\subfloat[]{\begin{tikzpicture}[baseline={([yshift=-0
cm]current bounding box.center)}, scale=1.4] 
	\node[] (a) at (0,0) {};
	\node[] (a1) at (0.5,0) {};
	\node[] (a2) at (1.5,0) {};
	\node[] (b) at (0,-0.5) {};
	\node[] (b1) at (1.5,-0.5) {};
	\node[] (c) at (0,-1) {};
	\node[] (c1) at (1,-1) {};
	\node[] (c2) at (1.5,-1) {};
	\node[] (p1) at ($(a)+(-0.2,0)$) {};
	\node[] (p2) at ($(c)+(-0.2,0)$) {};
	\node[] (p3) at ($(c2)+(0.2,0)$) {};
	\node[] (p4) at ($(a2)+(0.2,0)$) {};
	\draw[line width=0.15mm] (c.center) -- (a.center);
	\draw[line width=0.15mm] (b.center) -- (b1.center);
	\draw[line width=0.15mm] (c2.center) -- (a2.center);
	\draw[line width=0.15mm] (b1.center) -- (c1.center);
	\draw[line width=0.15mm] (b.center) -- (a1.center);
	\draw[line width=0.5mm] (p1.center) -- (p4.center);
	\draw[line width=0.5mm] (p2.center) -- (p3.center);
\end{tikzpicture}}
\caption{(a) Kinematics of the scattering process, exemplified by the tree-level diagram. 
The arrows indicate the direction of the momenta, thin and thick lines, respectively, denote the graviton and scalar matter propagators. (b) A Feynman diagram containing a Calabi-Yau three-fold.}
\label{fig: kinematics}
\end{figure}
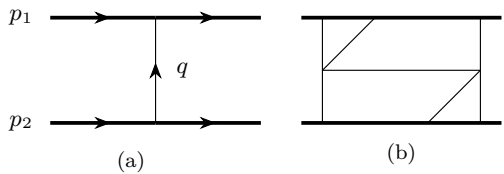

We study the scattering of two massive scalars with momenta $p_{i=1,2}$, masses $m_{i=1,2}$, and momentum transfer $q$; see fig.\ \ref{fig: kinematics}(a).
We decompose $p_1=\overline{p}_1{-}q/2$ and $p_2=\overline{p}_2{+}q/2$ into components orthogonal and along $q$~\cite{Landshoff:1969yyn,Parra-Martinez:2020dzs}, satisfying $\overline{p}_i \cdot q =0$. In \emph{Quantum} Field Theory, such a scattering process depends on the 
kinematic invariants $s=
(\overline{p}_1 {+} \overline{p}_2)^2$, $t=
q^2$, and $u=
(\overline{p}_1{-}\overline{p}_2)^2$ with $s+t+u=2m_1^2+2m_2^2$.

A \emph{classical} two-body problem, however, is further characterized by having a large angular momentum $J \gg \hbar$~\cite{Damour:2016gwp,Damour:2017zjx,Cheung:2018wkq,Bern:2019crd}. In natural units, this corresponds to $s, |u|, m_1^2, m_2^2 \sim  J^2|t| \gg |t| = |q|^2$, i.e., the classical limit corresponds to the limit of small $|q|$~\cite{Neill:2013wsa}. In order to focus on the classical dynamics, we will thus systematically implement this limit during the perturbative expansion. In practice, this is attained by a soft-$q$ expansion with the method of regions~\cite{Beneke:1997zp}, where hard (quantum) momenta $\sim m_i$ are suppressed in favor of soft momenta $\sim |q|$~\cite{Neill:2013wsa}. 

To identify the diagrams that contribute to the potential region, a power counting in  
$|q|$ is carried out within the soft expansion. 
Graviton propagators scale as $1/k^2 \sim |q|^{-2}$, where $k$ is the loop momentum, but the scalar matter propagators can be expanded and turn into linearized (or eikonal) propagators $\sim |q|^{-1}$. Introducing the soft four-velocities $u_i^\mu=\overline{p}_i^\mu /\overline{m}_i$, with the soft masses being $\overline{m}_i^2=\overline{p}_i^2=m_i^2-q^2/4$, so that $u_i^2=1$ and $u_i \cdot q=0$, the matter propagators become  
\begin{equation}
\frac{1}{(k+\overline{p}_i \pm \frac{q}{2})^2-m_i^2} = \frac{1}{m_i} \frac{1}{2 u_i \cdot k} + \mathcal{O}(q^2).
\end{equation}
In particular, this implies that the dependence on the masses $m_i$ factors out. 
Each loop integration measure scales as $|q|^4$, $n$-point graviton self-interaction vertices scale as $|q|^2G^{\frac{n}{2}-1}$,
 and the interaction of a matter line with $n$ gravitons scales as $|q|^0 G^{\frac{n}{2}}$. 
 
The leading term in the post-Minkowskian expansion (1PM) scales like $|q|^{-2}G$, as can be immediately seen from fig.\ \ref{fig: kinematics}(a). Then, each (loop) correction in the soft limit adds a factor of $|q|G$ \cite{Neill:2013wsa}. The classical contribution thus stems from diagrams that scale as 
$|q|^{L-2} G^{L+1}$ at $L$ loops~\cite{Neill:2013wsa}, corresponding to an $(L+1)$-PM correction. 
Diagrams with more powers of $|q|$, which are of quantum nature, become suppressed and can be discarded. Diagrams with fewer powers of $|q|$ are called superclassical (also known as iterations), which cancel when performing the matching of the full theory to the effective field theory~\cite{Bern:2019crd}; they can thus similarly be discarded.

Additionally, we will simplify the analysis by restricting ourselves to the (conservative) potential region within the soft expansion, where the momenta scale as $k^\mu=(\omega,\mathbf{k}) \sim |q|(v,1)$. 
In other words, radiation modes $k^\mu=(\omega,\mathbf{k}) \sim |q|(v,v)$ leading to radiation reaction and tail effects will also be discarded; see refs.~\cite{Bern:2021yeh,Dlapa:2022lmu,Herrmann:2021tct,Cho:2021arx,Kalin:2022hph} for a discussion.  

Since $q^2<0$ is the only remaining dimensionful scale in the classical limit, it can be fixed by dimensional analysis. Thus, the classical scattering process only depends non-trivially on $y = u_1 \cdot u_2$, often rewritten in terms of $y=\frac{x^2+1}{2x}$ to rationalize the square root $\sqrt{y^2-1}$ that regularly appears in the results.\footnote{Equivalently, it depends on the relative Lorentz factor $\sigma = p_1 \cdot p_2/(m_1 m_2) = (s-m_1^2-m_2^2)/(2\, m_1 m_2)=y + \mathcal{O}(q^2)$.}

\section{Detecting the geometry in Feynman integrals}
\label{sec: geometry}

Knowing which Feynman diagrams are relevant for the potential region in the classical limit, we now turn to the task of detecting the geometries in the corresponding Feynman integrals. We use two complementary methods: (i) differential equations and (ii) leading singularities computed via a loop-by-loop Baikov representation.

\paragraph{Picard-Fuchs operators} One frequently used approach to study the geometry in Feynman integrals is differential equations \cite{Kotikov:1990kg}. Feynman integrals can be reduced to so-called master integrals via integration-by-parts identities (IBPs)~\cite{Chetyrkin:1981qh}, as implemented, e.g., in {\texttt{\textup{FIRE}}}~\cite{Smirnov:2008iw,Smirnov:2023yhb}, {\texttt{\textup{KIRA}}}~\cite{Maierhofer:2017gsa}, and \texttt{LiteRed}~\cite{Lee:2012cn}.
Taking then a derivative of the vector of master integrals $\vec{I}$ with respect to one of the kinematic variables, e.g.\ $x$, yields a system of differential equations~\cite{Kotikov:1990kg} $\partial_x \vec{I} = A \, \vec{I}$, where the entries of the matrix $A$ depend on the dimension $D$ and the kinematics.

By taking further derivatives, this system of coupled first-order differential equations can be reduced to a single higher-order differential equation for an individual master integral $I_i$,
\begin{equation}
\mathcal{L}_n I_i = \text{inhom.}, \qquad \mathcal{L}_n=\frac{d^n}{d x^n} + \sum_{j=0}^{n-1} c_j(x) \frac{d^j}{d x^j}.
\end{equation}
The inhomogeneity stems from subsectors (also known as subtopologies), i.e., master integrals where a number of propagators are absent, corresponding to diagrams where those propagators are pinched. The so-called Picard-Fuchs operator $\mathcal{L}_n$, where $c_j(x)$ are rational functions, describes the corresponding homogeneous differential equation. If the rational factorization of the Picard-Fuchs operator produces an irreducible differential operator of order $r>1$, then this is a clear indication that the Feynman integral of interest is not polylogarithmic, but may involve a Calabi-Yau $(r{-}1)$-fold.\footnote{An alternative is for it to involve a higher-genus curve~\cite{Huang:2013kh,Marzucca:2023gto}. We can discriminate between these possibilities via a further analysis of $\mathcal{L}_n$ as well as the leading singularity.} The factorization of differential operators is implemented, e.g., in {\texttt{\textup{Maple}}}, and further studying the properties of the resulting irreducible differential operator uniquely identifies the geometry under consideration.

A Feynman integral inherits non-trivial geometries if its subsectors, and thus the inhomogeneity, involve them. However, we can remove the inhomogeneity~\cite{Primo:2016ebd} in the differential equation by considering the maximal cut, where all propagators $\frac{i}{Q_i^2-m_i^2}$
 are replaced by on-shell delta functions $\delta(Q_i^2-m_i^2)$. Since taking derivatives commutes with taking cuts, the operator $\mathcal{L}_n$ and thus the homogeneous differential equation yields the differential equation on the maximal cut.
As a consequence, we can look at the sectors one at a time for the purpose of detecting geometries.

\paragraph{Leading singularities via loop-by-loop Baikov} An alternative approach to detecting geometries in Feynman integrals is via leading singularities (LS), which are related to the maximally iterated discontinuity of the integral~\cite{Cachazo:2008vp}. Concretely, the leading singularity is obtained by taking the maximal cut as well as any further discontinuities. It can be calculated in the original momentum-space representation of the Feynman integral as well as in any parametric representation.
A Feynman integral is polylogarithmic if its leading singularity -- as well as the leading singularity of all its subsectors -- is algebraic. This allows us to analyze one subsector at a time, as in the case of differential equations.
If the leading singularity contains a non-trivial integral, this integral is indicative of the geometry contained in the Feynman integral.

Here, we will use the Baikov representation~\cite{Baikov:1996iu} for the calculation of the leading singularity. 
This representation can be obtained from the $D$-dimensional momentum-space representation by making a change of variables $z_i=Q_i^2-m_i^2$ so that the propagators characterizing the problem become the integration variables. For multi-loop problems, it is often necessary to introduce extra auxiliary propagators for the change of variables to be well defined. For a problem with $n_{\text{int}}$ propagators we need to add $\nISP=N_{\text{V}}-\nint$ extra variables accounting for irreducible scalar products (ISPs), where $N_{\text{V}} = \frac{1}{2} L(L+1) + E L$ is the number of independent scalar products that may be formed between the $L$ loop momenta and the $E = n_{\text{ext}}-1$ independent external momenta.

Overall, the Baikov parametrization reads, for the case where all propagators are raised to power one,
\begin{align}
\label{eq: Baikov}
I= \mathcal{G}^{\frac{-D+E+1}{2}} \int \frac{d^{N_{\text{V}}} z}{z_1 \cdots z_{\nint}} \ {\mathcal{B}(z)}^{\frac{D-L-E-1}{2}},
\end{align}
where we have dropped an overall factor that depends only on the space-time dimension $D$. Here, $\mathcal{B}(z)=\det G(k_1,\dots,k_L,p_1,\dots,p_E)$ is known as the Baikov polynomial, and $\mathcal{G} = \det G(p_1,\dots,p_E)$, where $G(Q_1,\dots,Q_n)$ denotes the Gram matrix, with entries $G_{ij}(Q_1,\dots,Q_n) = Q_i \cdot Q_j$.

The Baikov representation is particularly suitable for calculating the maximal cut, which is simply obtained by taking the residues at $z_i=0$ for $i=1,\dots, \nint$,
\begin{equation}
\label{eq: Baikov_max_cut}
I_{\text{max-cut}}\propto 
\mathcal{G}^{\frac{-D+E+1}{2}} \int d^\nISP z \ {\mathcal{B}(\underbrace{0,\dots,0}_{\nint},z)}^{\frac{D-L-E-1}{2}}.
\end{equation}
The leading singularity is then calculated by taking all possible further residues (if any) in the remaining $\nISP$ extra variables coming from the ISPs.

To facilitate the task of calculating the remaining residues, we use a slight variant of the Baikov representation, the so-called loop-by-loop Baikov representation~\cite{Frellesvig:2017aai}. It is derived by applying eq.~\eqref{eq: Baikov} separately to one loop at a time, which reduces the number of remaining integrals to $\nISP=L + \sum_{i=1}^L E_i-\nint$, often a considerable reduction. However, in doing so, the external momenta for each loop will also depend on other loop momenta; hence, the Gram determinant $\mathcal{G}$ will become a function $\mathcal{G}(z)$.\footnote{The complexity of the resulting representation moreover depends on the order in which the loop integrations are considered. Usually, the best choice of order starts with the sub-loops with the fewest edges; see ref.~\cite{Frellesvig:toapp} for details.}

Calculating the leading singularity via an integral representation of Feynman integrals has one important subtlety: non-trivial changes of variables may be required to expose all poles. To exclude the existence of additional changes of variables that expose further poles, we always use differential equations as cross-checks when finding non-trivial geometries via loop-by-loop Baikov. On the other hand, an algebraic leading singularity conclusively indicates the absence of non-trivial geometries.

\section{Relating and restricting geometries}
\label{sec: results1}

Before scanning over all relevant four-loop diagrams, we can use the leading singularity via the loop-by-loop Baikov representation to find general relations between the geometries in different diagrams, and show that whole classes of diagrams contain only trivial geometries, thus vastly reducing the number of diagrams we need to consider.
We present the corresponding derivations in ref.~\cite{Frellesvig:toapp} and only state the results here.

\begin{figure}[tb]
\centering
\subfloat[]{
$\LS \left( \begin{tikzpicture}[baseline={([yshift=-0.1cm]current bounding box.center)}, scale=0.9] 
	\node[] (a) at (0,0) {};
	\node[] (b) at (0,-1) {};
	\node[] (a1) at (1,0) {};
	\node[] (a2) at (1.75,0.2) {};
	\node[] (b1) at (1,-1) {};
	\node[] (b2) at (1.75,-1) {};
	\node[] (p1) at ($(a)+(-0.75,0.2)$) {};
	\node[] (p2) at ($(b)+(-0.75,0)$) {};
	\draw[line width=0.15mm] (b1.center) -- (a.center);
	\draw[line width=0.15mm] (b.center) -- (0.4,-0.6);
	\draw[line width=0.15mm] (0.6,-0.4) -- (a1.center);
	\draw[line width=0.5mm] (p1.center) -- (a2.center);
	\draw[line width=0.5mm] (p2.center) -- (b2.center);
	\fill[gray!50] (0.5,0.2) ellipse (0.7 and 0.3);
	\draw (0.5,0.2) ellipse (0.7 and 0.3);
	\node at (-0.15,-0.5)[circle,fill,inner sep=0.65pt]{};
	\node at (-0.35,-0.5)[circle,fill,inner sep=0.65pt]{};
	\node at (-0.55,-0.5)[circle,fill,inner sep=0.65pt]{};
	\node at (1.15,-0.5)[circle,fill,inner sep=0.65pt]{};
	\node at (1.35,-0.5)[circle,fill,inner sep=0.65pt]{};
	\node at (1.55,-0.5)[circle,fill,inner sep=0.65pt]{};
\end{tikzpicture} \right) \propto \LS \left(
\begin{tikzpicture}[baseline={([yshift=-0.1cm]current bounding box.center)}, scale=0.9]
	\node[] (a) at (0,0) {};
	\node[] (b) at (0,-1) {};
	\node[] (a1) at (1,0) {};
	\node[] (a2) at (2,0.2) {};
	\node[] (b1) at (1,-1) {};
	\node[] (b2) at (2,-1) {};
	\node[] (p1) at ($(a)+(-1,0.2)$) {};
	\node[] (p2) at ($(b)+(-1,0)$) {};
	\draw[line width=0.15mm] (b.center) -- (a.center);
	\draw[line width=0.15mm] (b1.center) -- (a1.center);
	\draw[line width=0.5mm] (p1.center) -- (a2.center);
	\draw[line width=0.5mm] (p2.center) -- (b2.center);
	\fill[gray!50] (0.5,0.2) ellipse (0.7 and 0.3);
	\draw (0.5,0.2) ellipse (0.7 and 0.3);
	\node at (-0.25,-0.5)[circle,fill,inner sep=0.65pt]{};
	\node at (-0.45,-0.5)[circle,fill,inner sep=0.65pt]{};
	\node at (-0.65,-0.5)[circle,fill,inner sep=0.65pt]{};
	\node at (1.25,-0.5)[circle,fill,inner sep=0.65pt]{};
	\node at (1.45,-0.5)[circle,fill,inner sep=0.65pt]{};
	\node at (1.65,-0.5)[circle,fill,inner sep=0.65pt]{};
\end{tikzpicture} \right)$}
\qquad
\subfloat[]{$\LS \left(
\begin{tikzpicture}[baseline={([yshift=-0.1cm]current bounding box.center)}, scale=0.9] 
	\node[] (a) at (-0.3,0) {};
	\node[] (a1) at (0.3,0) {};
	\node[] (a2) at (1,0) {};
	\node[] (a3) at (1.7,0) {};
	\node[] (a4) at (2.3,0) {};
	\node[] (b) at (0.3,-1) {};
	\node[] (b1) at (0.65,-1) {};
	\node[] (b2) at (1,-1) {};
	\node[] (b3) at (1.33,-1) {};
	\node[] (b4) at (1.7,-1) {};
	\node[] (c) at (-0.3,-1) {};
	\node[] (c1) at (2.3,-1) {};
	\node[] (t) at (1,-0.4) {};
	\node[] (t1) at (0.7,0) {};
	\node[] (t2) at (1.3,0) {};
	\node[] (p1) at ($(a)+(-0.3,0)$) {};
	\node[] (p2) at ($(c)+(-0.3,0)$) {};
	\node[] (p3) at ($(c1)+(0.3,0)$) {};
	\node[] (p4) at ($(a4)+(0.3,0)$) {};
	\draw[line width=0.15mm] (a.center) -- (b.center);
	\draw[line width=0.15mm] (a1.center) -- (b1.center);
	\draw[line width=0.15mm] (t.center) -- (b2.center);
	\draw[line width=0.15mm] (t.center) -- (t1.center);
	\draw[line width=0.15mm] (t.center) -- (t2.center);
	\draw[line width=0.15mm] (a3.center) -- (b3.center);
	\draw[line width=0.15mm] (a4.center) -- (b4.center);
	\draw[line width=0.5mm] (p1.center) -- (p4.center);
	\draw[line width=0.5mm] (p2.center) -- (p3.center);
	\fill[gray!50] (b2.center) ellipse (0.8 and 0.3);
	\draw (b2.center) ellipse (0.8 and 0.3);
	\node at (1.675,-0.36)[circle,fill,inner sep=0.65pt]{};
	\node at (1.8,-0.4)[circle,fill,inner sep=0.65pt]{};
	\node at (1.925,-0.44)[circle,fill,inner sep=0.65pt]{};
	\node at (0.325,-0.36)[circle,fill,inner sep=0.65pt]{};
	\node at (0.2,-0.4)[circle,fill,inner sep=0.65pt]{};
	\node at (0.075,-0.44)[circle,fill,inner sep=0.65pt]{};
\end{tikzpicture} \right) = 0$}
\caption{Simplifying relations for the geometries in PM Feynman integrals: (a) The vertices at matter lines become effectively orderless; (b) Dangling triangles have vanishing leading singularities.}
\label{fig: simplification_geometries}
\end{figure}
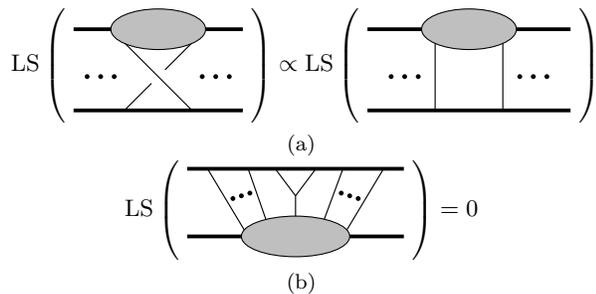

\paragraph{Planar and non-planar diagrams} For the purpose of PM Feynman integral geometries, vertices at scalar lines are effectively orderless, see fig.\ \ref{fig: simplification_geometries}(a), which allows us to relate the leading singularity of non-planar diagrams to planar counterparts.\footnote{Starting at four loops, however, there also exist diagrams whose non-planarity does not originate from the order of the vertices, and cannot be related to planar counterparts.}

\paragraph{Dangling triangles} Triangles at a matter line with only cubic vertices such as in fig.\ \ref{fig: simplification_geometries}(b) -- which we call dangling triangles -- have vanishing leading singularity. Thus, they are fully reducible to subsectors.

\begin{figure}[tb]
\centering
\begin{tikzpicture}[baseline={([yshift=-0.1cm]current bounding box.center)},scale=0.68] 
	\node[] (a) at (0,0) {};
	\node[] (a1) at (1,0) {};
	\node[] (a2) at (2,0) {};
	\node[] (a3) at (3,0) {};
	\node[] (a4) at (4,0) {};
	\node[] (a5) at (5,0) {};
	\node[] (a6) at (6,0) {};
	\node[] (b) at (0,-1) {};
	\node[] (b1) at (1,-1) {};
	\node[] (b2) at (2,-1) {};
	\node[] (b3) at (3,-1) {};
	\node[] (b4) at (4,-1) {};
	\node[] (b5) at (5,-1) {};
	\node[] (b6) at (6,-1) {};
	\node[] (c) at (0,-2) {};
	\node[] (c1) at (1,-2) {};
	\node[] (c2) at (2,-2) {};
	\node[] (c3) at (3,-2) {};
	\node[] (c4) at (4,-2) {};
	\node[] (c5) at (5,-2) {};
	\node[] (c6) at (6,-2) {};
	\node[] (p1) at ($(a)+(-0.3,0)$) {};
	\node[] (p2) at ($(c)+(-0.3,0)$) {};
	\node[] (p3) at ($(c6)+(0.3,0)$) {};
	\node[] (p4) at ($(a6)+(0.3,0)$) {};
	\draw[line width=0.15mm] (b.center) -- (a.center);
	\draw[line width=0.15mm] (b1.center) -- (a1.center);
	\draw[line width=0.15mm] (b3.center) -- (a3.center);
	\draw[line width=0.15mm] (b4.center) -- (a4.center);
	\draw[line width=0.15mm] (b6.center) -- (a6.center);
	\draw[line width=0.15mm] (b.center) -- (b6.center);
	\draw[line width=0.15mm] (b.center) -- (c.center);
	\draw[line width=0.15mm] (b2.center) -- (c2.center);
	\draw[line width=0.15mm] (b5.center) -- (c5.center);
	\draw[line width=0.15mm] (b6.center) -- (c6.center);
	\draw[line width=0.5mm] (p1.center) -- (p4.center);
	\draw[line width=0.5mm] (p2.center) -- (p3.center);
	\node at (5,-0.5)[circle,fill,inner sep=0.7pt]{};
	\node at (5.25,-0.5)[circle,fill,inner sep=0.7pt]{};
	\node at (4.75,-0.5)[circle,fill,inner sep=0.7pt]{};
	\node at (5.5,-1.5)[circle,fill,inner sep=0.7pt]{};
	\node at (5.25,-1.5)[circle,fill,inner sep=0.7pt]{};
	\node at (5.75,-1.5)[circle,fill,inner sep=0.7pt]{};
\end{tikzpicture}
\caption{The Mondrian family of Feynman diagrams.
}
\label{fig: diagrams_geometries}
\end{figure}
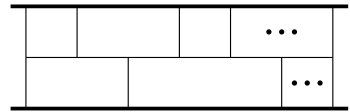

With this knowledge, we can limit the study of the geometries to diagrams of the Mondrian\footnote{This notation goes back to the resemblance to paintings by Piet Mondrian, see e.g.\  ref.~\cite{Bern:2004kq}.} family of fig.~\ref{fig: diagrams_geometries}, to their non-planar variations that are not related by reordering vertices at the scalar lines, as well as to their sub-topologies.

\section{A Calabi-Yau three-fold at $\mathbf{\mathcal{O}(G^5)}$}
\label{sec: results2}

The vast majority of the diagrams at four loops 
have algebraic leading singularities, indicating that they are polylogarithmic. Besides, we encounter several diagrams that (more or less trivially) contain the same K3 surface that was identified at three loops.
We will present the full results of our systematic analysis in ref.~\cite{Frellesvig:toapp}. Here, we highlight one particular new geometry that we identify: 
a Calabi-Yau three-fold. 

Let us consider the scalar diagram depicted in fig.~\ref{fig: kinematics}(b), a subsector of the four-loop Mondrian diagram that is obtained by pinching two propagators. It has the following leading singularity in $D=4$:
\begin{equation}
\label{eq: LS_diag_Calabi_Yau}
\LS \left( \begin{tikzpicture}[baseline={([yshift=-0.1cm]current bounding box.center)}, scale=0.75] 
	\node[] (a) at (0,0) {};
	\node[] (a1) at (0.5,0) {};
	\node[] (a2) at (1,0) {};
	\node[] (b) at (0,-0.5) {};
	\node[] (b1) at (1,-0.5) {};
	\node[] (c) at (0,-1) {};
	\node[] (c1) at (0.5,-1) {};
	\node[] (c2) at (1,-1) {};
	\node[] (p1) at ($(a)+(-0.2,0)$) {};
	\node[] (p2) at ($(c)+(-0.2,0)$) {};
	\node[] (p3) at ($(c2)+(0.2,0)$) {};
	\node[] (p4) at ($(a2)+(0.2,0)$) {};
	\draw[line width=0.15mm] (c.center) -- (a.center);
	\draw[line width=0.15mm] (b.center) -- (b1.center);
	\draw[line width=0.15mm] (c2.center) -- (a2.center);
	\draw[line width=0.15mm] (b1.center) -- (c1.center);
	\draw[line width=0.15mm] (b.center) -- (a1.center);
	\draw[line width=0.5mm] (p1.center) -- (p4.center);
	\draw[line width=0.5mm] (p2.center) -- (p3.center);
\end{tikzpicture} \right) \propto  x \int \frac{d t_1 d t_2 d t_3}{\sqrt{P_8(t_1,t_2,t_3)}}
\end{equation}
with
\begin{align}
\nonumber
P_8(t_1, t_2, t_3)  = {}& (1{+}t_1)^2 \, (1{+}t_2)^2 \, (t_1 t_2 {+} t_3^2)^2 \, (1 {-} x^2)^2 \\ 
&+{} 64 \, t_1^2 \, t_2^2 \, (1{+}t_1{+}t_2) \, t_3^2 \, x^2,
\label{eq: P8}
\end{align}
where we have set $q^2=-1$ since the dependence on $q^2$ can always be recovered via dimensional analysis;
see appendix~\ref{sec: Baikov_Calabi_Yau}
for details of the calculation.
It contains the square root of a polynomial $P_8(t_1,t_2,t_3)$ of degree 8 in three variables.
Homogenizing this polynomial to $\widetilde{P}_8(t_1,t_2,t_3,t_4)=P_8(t_1/t_4,t_2/t_4,t_3/t_4)\, t_4^8$,  we obtain a homogeneous equation of degree 8 that defines a co-dimension-one hypersurface in weighted projective space $[t_1,t_2,t_3,t_4,t_5]\sim [\lambda^1 t_1,\lambda^1 t_2, \lambda^1 t_3, \lambda^1 t_4,\lambda^4 t_5]\in\mathbb{WP}^{1,1,1,1,4}$,
\begin{equation}
 t_5^2-\widetilde{P}_8(t_1,t_2,t_3,t_4)=0.
\end{equation}
Since the degree of the equation is equal to the sum of the weights, its solution generically defines a Calabi-Yau three-fold \cite{Hubsch:1992nu,Bourjaily:2019hmc}.

From the perspective of the differential equation, we find an irreducible Picard-Fuchs operator of order $4$ in $D=4$,
\begin{align}
\label{eq: Picard_Fuchs_Calabi_Yau}
\mathcal{L}_4 = &\, 
 \frac{\partial^4}{\partial x^4} + \frac{2 - 16 x^2 - 10 x^4}{x(1-x^4)} \frac{\partial^3}{\partial x^3} \nonumber \\
& \, + \frac{1 - 28 x^2 + 46 x^4 + 68 x^6 + 25 x^8}{x^2 (1 - x^4)^2} \frac{\partial^2}{\partial x^2} \nonumber \\
& \, - \frac{1 + 11 x^2 - 54 x^4 + 22 x^6 + 37 x^8 + 15 x^{10}}{x^3 (1 - x^2)^3 (1 + x^2)^2} \frac{\partial}{\partial x} \nonumber \\
& \, + \frac{1 + 3 x^2 + 20 x^4 + 3 x^6 + x^8}{x^4 (1 - x^4)^2},
\end{align}
further confirming the Calabi-Yau geometry found with the leading singularity. Moreover, $\mathcal{L}_4$ satisfies all conditions for being a Calabi-Yau operator~\cite{Bogner:2013kvr,Morrison:1991cd,Ceresole:1992su,Almkvist:2004kj,Almkvist:2005qoo,Yang:2008,Bogner:2011,vanStraten:2017,Candelas:2021tqt}, ultimately confirming that the geometry we found is a Calabi-Yau three-fold~\cite{Brammer:toapp}. We compute the solution to the corresponding differential equation in $D=4-2\varepsilon$ in upcoming work~\cite{Brammer:toapp}. 

Moreover, this 
new geometry within the PM expansion will appear in all diagrams containing fig.~\ref{fig: kinematics}(b) as a subsector, as well as in those related to it by the equivalences of fig.~\ref{fig: simplification_geometries}. While the full calculation of the potential at four-loop order is beyond the scope of this work, no full cancellation of geometries in the sum of diagrams has been observed at three-loop order, strongly suggesting that the corresponding functions also occur in the result at four-loop order.

\section{Conclusion and Outlook}
\label{sec: conclusion}

In this letter, we have studied Feynman integral geometries that contribute to the classical conservative dynamics of black holes in the post-Minkowskian expansion and thus to the emission of gravitational waves in the inspiraling phase of black hole mergers.
We demonstrated that leading singularities via loop-by-loop Baikov, complemented by differential equations on the maximal cut, provide a highly efficient method for detecting these geometries.
In particular, we identify -- for the first time for gravitational waves -- a Calabi-Yau three-fold at four loops, i.e, at fifth post-Minkowskian order. While the third-order differential equation of univariate K3 surfaces is the symmetric square of a second-order differential equation~\cite{Joyce:1972,Joyce:1973,Verrill:1996,Doran:1998hm} and can thus be solved in terms of known functions; the fourth-order differential equation of our Calabi-Yau three-fold is not a symmetric power \cite{Brammer:toapp}, indicating that completely new functions arise at fifth post-Minkowskian order!

In upcoming work \cite{Frellesvig:toapp}, we will use the methods described here to fully classify the Feynman integral geometries that occur through fifth post-Minkowskian order.

In a second upcoming work \cite{Brammer:toapp}, we will calculate the Feynman integral involving the newly identified Calabi-Yau three-fold via its differential equation.
This can be achieved by bringing the differential equation into the so-called $\varepsilon$-factorized form, which was recently generalized from polylogarithms \cite{Henn:2013pwa} to elliptic functions and Calabi-Yau integrals \cite{Frellesvig:2021hkr, Pogel:2022yat, Pogel:2022vat, Gorges:2023zgv, Frellesvig:2023iwr}.

One immediate question for future work is whether the K3 surface at three loops and the Calabi-Yau three-fold at four loops are part of a family of Calabi-Yau $(L-1)$-folds at $L$-loop order, analogously to previously identified integral families~\cite{Broadhurst:1993mw,Bourjaily:2018ycu,Bourjaily:2018yfy,Bonisch:2021yfw,Broedel:2021zij,Duhr:2022pch,Lairez:2022zkj,Pogel:2022vat,Duhr:2022dxb,Cao:2023tpx}.  
Moreover, it would be interesting to systematically use similar techniques at higher loops, for radiation reaction \cite{Bern:2021yeh,Dlapa:2022lmu,Herrmann:2021tct,Cho:2021arx,Kalin:2022hph} as well as for the waveform \cite{Brandhuber:2023hhy,Herderschee:2023fxh,Elkhidir:2023dco,Georgoudis:2023lgf}.

\begin{acknowledgments}

We are indebted to Zvi Bern, Enrico Herrmann, and Michael Ruf for very fruitful discussions. We thank Sebastian Pögel for collaboration on a related project~\cite{Brammer:toapp}, as well as Zvi Bern and Sebastian Pögel for comments on the draft. We moreover thank Christoph Dlapa, Matt von Hippel, Zhengwen Liu, Andres Luna, Andrew McLeod, Jan Plefka, and Radu Roiban for interesting discussions and Alexander Smirnov for helpful communication. We thank Jan Plefka for making us aware of ongoing parallel work~\cite{Plefka:comm}. This work was supported by the research grant 00025445 from Villum Fonden.
H.F.\ is moreover supported by a Carlsberg Foundation Reintegration Fellowship and has received funding from the European Union's Horizon 2020 research and innovation program under the Marie Sk{\l}odowska-Curie grant agreement No. 847523 `INTERACTIONS'.
R.M. is also grateful for the hospitality and support received from the Mani L. Bhaumik Institute for Theoretical Physics during a crucial stage of this project.

\end{acknowledgments}

\begin{appendix}
\section{Loop-by-loop Baikov analysis}
\label{sec: Baikov_Calabi_Yau}

In this appendix, we calculate the leading singularity of the diagram depicted in fig.~\ref{fig: kinematics}(b) via a loop-by-loop Baikov parametrization, finding that it contains a Calabi-Yau three-fold.

We parametrize the diagram as follows,
\begin{equation}
\label{eq: diag_Calabi_Yau}
\begin{tikzpicture}[baseline=(current bounding box.center), scale=0.6] 
	\node[] (a) at (0,0) {};
	\node[] (a1) at (2,0) {};
	\node[] (a2) at (6,0) {};
	\node[] (b) at (0,-2) {};
	\node[] (b1) at (6,-2) {};
	\node[] (c) at (0,-4) {};
	\node[] (c1) at (4,-4) {};
	\node[] (c2) at (6,-4) {};
	\node[label=left:{$\overline{p}_1{-}\frac{q}{2}$}] (p1) at ($(a)+(-1,0)$) {};
	\node[label=left:{$\overline{p}_2{+}\frac{q}{2}$}] (p2) at ($(c)+(-1,0)$) {};
	\node[label=right:{$\overline{p}_2{-}\frac{q}{2}$}] (p3) at ($(c2)+(1,0)$) {};
	\node[label=right:{$\overline{p}_1{+}\frac{q}{2}$}] (p4) at ($(a2)+(1,0)$) {};
	\draw[line width=0.15mm, postaction={decorate}] (b.center) -- node[sloped, allow upside down, label={[xshift=0.15cm, yshift=0cm]$k_1$}] {\midarrow} (a.center);
	\draw[line width=0.15mm, postaction={decorate}] (a1.center) -- node[sloped, allow upside down, label={[xshift=-0.05cm, yshift=0.45cm]$k_1{-}k_2$}] {\midarrow} (b.center);
	\draw[line width=0.15mm, postaction={decorate}] (b1.center) -- node[sloped, allow upside down, label={[xshift=-1.6cm, yshift=0.45cm]$k_4{-}k_3$}] {\midarrow} (c1.center);
	\draw[line width=0.15mm, postaction={decorate}] (a2.center) -- node[sloped, allow upside down, label={[xshift=-0.15cm, yshift=0cm]$k_2{-}q$}] {\midarrow} (b1.center);
	\draw[line width=0.15mm, postaction={decorate}] (b1.center) -- node[sloped, allow upside down, label={[xshift=0cm, yshift=0.8cm]$k_2{+}k_3$}] {\midarrow} (b.center);
	\draw[line width=0.15mm, postaction={decorate}] (b.center) -- node[sloped, allow upside down, label={[xshift=-0.9cm, yshift=0.05cm]$k_3$}] {\midarrow} (c.center);
	\draw[line width=0.15mm, postaction={decorate}] (c2.center) -- node[sloped, allow upside down, label={[xshift=1.4cm, yshift=0cm]$k_4{+}q$}] {\midarrow} (b1.center);
	\draw[line width=0.5mm, postaction={decorate}] (a.center) -- node[sloped, allow upside down, label={[xshift=0cm, yshift=-0.15cm]$2u_1\cdot k_1$}] {\midarrow} (a1.center);
	\draw[line width=0.5mm, postaction={decorate}] (a1.center) -- node[sloped, allow upside down, label={[xshift=0cm, yshift=-0.15cm]$2u_1\cdot k_2$}] {\midarrow} (a2.center);
	\draw[line width=0.5mm, postaction={decorate}] (c.center) -- node[sloped, allow upside down, label={[xshift=0cm, yshift=-0.85cm]$2u_2\cdot k_3$}] {\midarrow} (c1.center);
	\draw[line width=0.5mm, postaction={decorate}] (c1.center) -- node[sloped, allow upside down, label={[xshift=0cm, yshift=-0.85cm]$2u_2\cdot k_4$}] {\midarrow} (c2.center);
	\draw[line width=0.5mm, postaction={decorate}] (p1.center) -- node[sloped, allow upside down] {\midarrow} (a.center);
	\draw[line width=0.5mm, postaction={decorate}] (a2.center) -- node[sloped, allow upside down] {\midarrow} (p4.center);
	\draw[line width=0.5mm, postaction={decorate}] (p2.center) -- node[sloped, allow upside down] {\midarrow} (c.center);
	\draw[line width=0.5mm, postaction={decorate}] (c2.center) -- node[sloped, allow upside down] {\midarrow} (p3.center);
\end{tikzpicture}.
\end{equation}
Taking the loop ordering $k_1\to k_4 \to k_3 \to k_2$ and defining the extra Baikov variables $z_{12}=(k_3{+}q)^2$, $z_{13}=k_2^2$, and $z_{14}=2u_2\cdot k_2$, we have
\begin{align}
 \begin{tikzpicture}[baseline={([yshift=-0.1cm]current bounding box.center)}, scale=0.75] 
	\node[] (a) at (0,0) {};
	\node[] (a1) at (0.5,0) {};
	\node[] (a2) at (1,0) {};
	\node[] (b) at (0,-0.5) {};
	\node[] (b1) at (1,-0.5) {};
	\node[] (c) at (0,-1) {};
	\node[] (c1) at (0.5,-1) {};
	\node[] (c2) at (1,-1) {};
	\node[] (p1) at ($(a)+(-0.2,0)$) {};
	\node[] (p2) at ($(c)+(-0.2,0)$) {};
	\node[] (p3) at ($(c2)+(0.2,0)$) {};
	\node[] (p4) at ($(a2)+(0.2,0)$) {};
	\draw[line width=0.15mm] (c.center) -- (a.center);
	\draw[line width=0.15mm] (b.center) -- (b1.center);
	\draw[line width=0.15mm] (c2.center) -- (a2.center);
	\draw[line width=0.15mm] (b1.center) -- (c1.center);
	\draw[line width=0.15mm] (b.center) -- (a1.center);
	\draw[line width=0.5mm] (p1.center) -- (p4.center);
	\draw[line width=0.5mm] (p2.center) -- (p3.center);
\end{tikzpicture} &\propto \int \frac{d z_1 \cdots d z_{14}}{z_1 \cdots z_{11} \, \sqrt{\det G(k_2,u_1)} \sqrt{\det G(k_3{+}q,u_2)}} \nonumber \,\,\,\,\\
& \!\!\!\!\!\!\!\!\!\!\!\!\!\!\!\! \times \frac{1}{\sqrt{\det G(k_3,k_2,u_2,q)} \sqrt{\det G(k_2,u_1,u_2,q)}},
\end{align}
and 
\begin{align}
& \LS \left( \begin{tikzpicture}[baseline={([yshift=-0.1cm]current bounding box.center)}, scale=0.75] 
	\node[] (a) at (0,0) {};
	\node[] (a1) at (0.5,0) {};
	\node[] (a2) at (1,0) {};
	\node[] (b) at (0,-0.5) {};
	\node[] (b1) at (1,-0.5) {};
	\node[] (c) at (0,-1) {};
	\node[] (c1) at (0.5,-1) {};
	\node[] (c2) at (1,-1) {};
	\node[] (p1) at ($(a)+(-0.2,0)$) {};
	\node[] (p2) at ($(c)+(-0.2,0)$) {};
	\node[] (p3) at ($(c2)+(0.2,0)$) {};
	\node[] (p4) at ($(a2)+(0.2,0)$) {};
	\draw[line width=0.15mm] (c.center) -- (a.center);
	\draw[line width=0.15mm] (b.center) -- (b1.center);
	\draw[line width=0.15mm] (c2.center) -- (a2.center);
	\draw[line width=0.15mm] (b1.center) -- (c1.center);
	\draw[line width=0.15mm] (b.center) -- (a1.center);
	\draw[line width=0.5mm] (p1.center) -- (p4.center);
	\draw[line width=0.5mm] (p2.center) -- (p3.center);
\end{tikzpicture} \right) \propto  
\int \frac{x \, d z_{12} d z_{13} d z_{14}}{\sqrt{(x^2-1)^2 (1+z_{13})^2 +4x^2 z_{14}^2}}  \nonumber \\
&   
\,\,\times \frac{1}{\sqrt{z_{12}} \sqrt{z_{13}}\sqrt{z_{14}^2 (1+z_{12})^2 - 4 z_{12} z_{13}(z_{12} + z_{13} +1)}} 
,
\end{align}
where we already set $q^2=-1$ since the dependence on it can be recovered by dimensional analysis. We can now introduce $t_{14}$ via \begin{equation}
z_{14} = r_1-\frac{(r_2-r_1)(1-t_{14})^2}{4t_{14}},
\end{equation}
where $r_{1,2}$ are the roots of the quadratic polynomial with respect to $z_{14}$, to rationalize the first 
square root. Subsequently rescaling $t_{14} \to t_{14}/(\sqrt{z_{12}} \sqrt{z_{13}})$, and relabeling $z_{12} \to t_1$, $z_{13} \to t_2$ and $t_{14} \to t_3$, we obtain eq.~\eqref{eq: LS_diag_Calabi_Yau}.

\end{appendix}

\bibliography{References}

\end{document}